\documentclass[aps,prc,letterpaper,11pt,amsmath,amssymb,twoside,nofootinbib,showpacs,preprint]{revtex4}
\usepackage{graphicx}
\usepackage{amsmath}
\usepackage{amsfonts}
\usepackage{amssymb}

\begin{document}
\newcommand{\be}{\begin{equation}}
\newcommand{\ee}{\end{equation}}
\newcommand{\bea}{\begin{eqnarray}}
\newcommand{\eea}{\end{eqnarray}}
\newcommand{\bef}{\begin{figure}}
\newcommand{\eef}{\end{figure}}
\newcommand{\bce}{\begin{center}}
\newcommand{\ece}{\end{center}}
\title{A new perspective on the Faddeev equations and the $\bar{K}NN$ system from chiral dynamics and unitarity in coupled channels.}

\author{E. Oset$^1$, D. Jido$^2$, T. Sekihara$^3$, A. Martinez Torres$^2$, K.P. Khemchandani$^3$, M. Bayar$^{1,5}$ and J. Yamagata-Sekihara$^1$}
\affiliation{$^1$Instituto de F{\'\i}sica Corpuscular (centro mixto CSIC-UV)\\
Institutos de Investigaci\'on de Paterna, Aptdo. 22085, 46071, Valencia, Spain\\
$^2$Yukawa Institute for Theoretical Physics, 
Kyoto University, Kyoto 606-8502, Japan\\
$^3$Department of Physics, Tokyo Institute of Technology, 
Tokyo 152-8551, Japan\\
$^4$Research Center for Nuclear Physics (RCNP), Osaka University, Ibaraki, Osaka 567-0047, Japan\\
$^5$Department of Physics, Kocaeli University, 41380 Izmit, Turkey}

\begin{abstract}

 We review recent work concerning the $\bar{K}N$ interaction and Faddeev equations with chiral dynamics which allow us to look at the $\bar{K}NN$ from a different perspective and pay attention to problems that have been posed in previous studies on the subject. We show results which provide extra experimental evidence on the existence of two $\Lambda(1405)$ states. We then show the findings of a recent approach to Faddeev equations using chiral unitary dynamics, where an explicit cancellation of the two body off shell amplitude with three body forces stemming from the same chiral Lagrangians takes place. This removal of the unphysical off shell part of the amplitudes is most welcome and renders the approach unambiguous, showing that only on shell two body amplitudes need to be used. With this information in mind we use an approximation to the Faddeev equations within the fixed center approximation to study the  $\bar{K}NN$ system, providing answers within this approximation to questions that have been brought before and evaluating binding energies and widths of this three body system.  As a novelty with respect to recent work on the topic we find a bound state of the system with spin S=1, like a bound state of $\bar{K}$-deuteron, less bound that the one of  S=0, where all recent efforts have been devoted. The width is relatively large in this case, suggesting problems in a possible experimental observation. 
\end{abstract}


\vspace{1cm}

\date{\today}

\maketitle

\section{Introduction}
 \label{sec:intro}
 
 The $\bar{K}N$ interaction has received much attention in the Literature with work done as early as in \cite{Dalitz:1960du,Dalitz:1967fp,Veit:1984an}. The advent of chiral dynamics in its unitarized form, the so called chiral unitary approach, has brought a new perspective to the problem and showed the importance of coupled channels and unitarity \cite{Kaiser:1995eg,angels,ollerulf,Lutz:2001yb,Oset:2001cn,Hyodo:2002pk,Jido:2003cb,Borasoy:2005ie,Oller:2006jw,Borasoy:2006sr}.
 One of the novel aspects of these works has been the finding of two poles, and two states, rather than one, associated to the experimental peak of the $\Lambda(1405)$ resonance. It was found in \cite{Jido:2003cb} that the two poles stem from a SU(3) singlet and octet of the chiral dynamical theory. One pole is around 1390 MeV, is wide ($\Gamma>100$ MeV) and couples mostly to $\pi \Sigma$. The other pole appears in all these theories around 1420 MeV, is narrow ($\Gamma=30$ MeV) and couples mostly $\bar{K}N$.  Due to the existence of these two poles the peak observed in experiment should be different in different reactions, as has been the case in the reactions $K^- p \to \pi^0\pi^0 \Sigma^0$ \cite{Prakhov:2004an} and $K^- d \to n \pi \Sigma$ \cite{Braun:1977wd}, which have been analyzed in 
 \cite{Magas:2005vu} and \cite{Jido:2009jf} respectively, and from where evidence was found for a resonance around 1420 MeV.

 In a recent paper \cite{Jido:2009jf} we reported on the interest of the $K^{-} d \to \pi \Sigma n$ reaction, which was measured in \cite{Braun:1977wd}. The idea is that the $K^{-}$ scatters with a neutron, loses energy and can interact with the proton to produce the $\Lambda(1405)$, which was seen in this reaction in the $\pi \Sigma$ spectrum. In this way the production of the resonance is induced by a $K^{-}$ and this guarantees that the state excited if mostly the one appearing at higher energy and narrow, out of the two $\Lambda(1405)$ states found in \cite{Jido:2003cb}. One of the conditions for the success of the experiment was to use kaons in flight. The reason is that the process that shows clearly the resonance peak is the double scattering. If the kaon is away from threshold, the dominant one body scattering is far away of the resonance region and peaks at higher energies. At lower energies of the kaon, the peak of the one body collision appears close to threshold and blurs the signal of the $\Lambda(1405)$ coming from double scattering. More recently we have also shown \cite{sekideu} that with the DAFNE conditions, where the kaons come with low energy from the decay of the $\phi$,
one can still see the good signal for the resonance, but on the condition that neutrons are detected simultaneously in the forward direction in the CM, which drastically reduces the background from single scattering.

Since we are working with the $K^-d$ reaction, we here revise our recent
results for the $K^-d$ system from Ref. \cite{melahat}, finding a weakly bound (but broad)
$S=1$ state. This is done by noting that the use of chiral dynamics in the Faddeev equations has shown that the off shell part of the two body amplitudes, which appears in the Faddeev equations and is known to be unphysical, gets cancelled by three body contact terms stemming from the same chiral Lagrangians \cite{MartinezTorres:2007sr,MartinezTorres:2008gy,Khemchandani:2008rk}. In this way the unphysical part of the amplitudes is removed and only the on shell two body amplitudes are needed in this approach. These off shell effects are responsible for the differences in the three body  calculations that use  input potentials producing the same on shell two body amplitudes. The new method removes these ambiguities and relies upon physical on shell amplitudes. From these perspectives we present here new results for the three body system $\bar{K} NN$ ($S=1$).

The plan of the paper is as follows: in Section II, we show the formalism for the $K^{-} d \to \pi \Sigma n$ reaction and
 in Sections III and IV, we summarize the new chiral approach to Faddeev equations. Section V outlines the  $\bar{K}NN$ system while Section VI deals with the fixed center formalism for the $\bar{K}NN$ 
($S=1$) interaction. In section VII we show the result of $\bar{K}NN$ ($S=1$) system taking into account charge exchange processes.
In Section VIII we calculate explicitly the $\pi\Sigma N$ channel in three body system and we show the result for this process. 
In Section IX, we present the result for the $K^-d $ scattering length and we compare our result with other results in the  Literature and some calculations are presented in Section X.

\section{Formalism for the $K^{-} d \to \pi \Sigma n$ reaction}

\begin{figure}
\begin{center}
\centerline{\includegraphics[width=8.5cm]{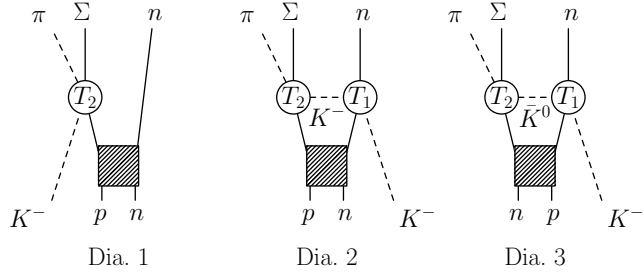}}
\caption{Diagrams for the calculation of the $K^{-}d \to \pi\Sigma n$ reaction.
$T_{1}$ and $T_{2}$ denote the scattering amplitudes for $\bar KN \to \bar KN$
and $\bar K N \to \pi \Sigma$, respectively.  \label{fig2jido}}
\end{center}
\end{figure}

The ${\cal T}$  matrix for the $K^{-} d \to \pi \Sigma n$ reaction is given, using single and double scattering terms, by the sum of the contribution of the three diagrams of  Fig.~\ref{fig2jido}, ${\cal T} =  {\cal T}_{1} + {\cal T}_{2} + {\cal T}_{3}$,
where the different amplitudes are given by 

\begin{equation}
  {\cal T}_{1} = T_{K^{-}p \rightarrow \pi \Sigma}(M_{\pi\Sigma}) \, \tilde{\varphi}(\vec p_{n} - \frac{\vec p_{d}}{2}). \label{eq:T1}
\end{equation}

\begin{eqnarray}
  {\cal T}_{2}& =&  T_{K^{-}p \rightarrow \pi \Sigma}(M_{\pi\Sigma}) 
 \int \frac{d^{3}q}{(2\pi)^{3}} \frac{\tilde \varphi (\vec q+\vec p_{n}-\vec k - \frac{\vec p_{d}}{2})}{q^{2}-m_{K}^{2} + i\epsilon}  
 T_{K^{-}n \rightarrow K^{-}n}(W_{1}) \ . \label{eq:T2}
\end{eqnarray}

\begin{eqnarray}
  {\cal T}_{3}& =& - T_{\bar K^{0}n \rightarrow \pi \Sigma}(M_{\pi\Sigma}) 
  \int \frac{d^{3}q}{(2\pi)^{3}} \frac{\tilde \varphi (\vec q+\vec p_{n}-\vec k - \frac{\vec p_{d}}{2})}{q^{2}-m_{K}^{2} + i\epsilon}
  T_{K^{-}p \rightarrow \bar K^{0}n}(W_{1})  \ . \label{eq:T3}
\end{eqnarray}
with $\tilde{\varphi}(\vec p_{n} - \frac{\vec p_{d}}{2})$ the deuteron wave function in momentum space and 
\begin{eqnarray}
   q^{0} &=& M_{N} + k^{0} - p_{n}^{0}\ , \\
   W_{1} &=& \sqrt{(q^{0}+p_{n}^{0})^{2}-(\vec q + \vec p_{n})^{2}} \ .
\end{eqnarray}
For $q^0$ we have assumed that the deuteron at rest has energy 
$2M_N -B$, and we have taken half of it for one nucleon, 
neglecting the small binding energy. The variable $W_1$ depends, 
however, on the running $\vec{q}$ variable.

\begin{figure}
\begin{center}
%
\includegraphics[width=8cm]{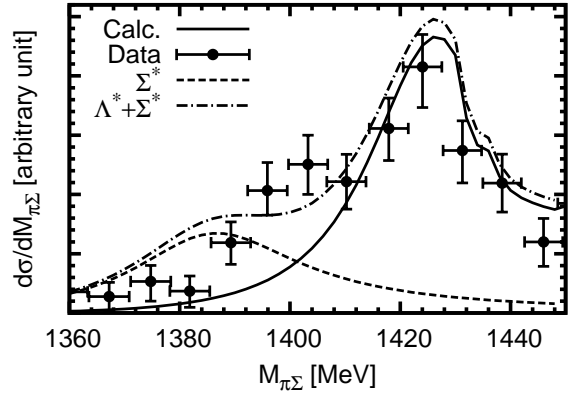}
%
\caption{$\pi\Sigma$ invariant mass spectra of $K^{-}d \to \pi^{+}\Sigma^{-}n$
with 800 MeV/c of incident $K^{-}$ momentum. Data from Ref. \cite{Braun:1977wd}. Calculations from Ref. \cite{Jido:2009jf}. 
}
\label{figyeni}
\end{center}
\end{figure}

 In fig. \ref{figyeni} we show the results that we obtain in arbitrary units, although we have found that there is also agreement  with the integrated cross section \cite{Jido:2009jf}. One observes experimentally a peak around 1420 MeV, as one would expect since the resonance excitation is induced by the $K^{-}p $ and the 
 $K^{-}p$ couples mostly to the $\Lambda^{*}$ state at 1420 MeV. The discrepancy of the theory with experiment in the low energy region is solved by adding a contribution from the excitation of the $\Sigma^{*}(1385)$ using information provided by experiment. The position of the experimental peak and the theoretical explanation for it provide a strong support for the existence of the two poles predicted by the chiral unitary approach.

\section{New chiral approach to Faddeev equations}

Our understanding of baryon resonances is undergoing continuous change. From the classical picture of the baryons made out of three constituent quarks, passing through attempts to represent some of them in terms of pentaquarks, to the more recent description of some of them in terms of meson baryon molecules. In this sense, the low lying $J^P= 1/2^-$ resonances, $\Lambda(1405)$, $\Lambda(1670)$, etc. can be represented as  composite states of meson and baryon, dynamically generated from the meson baryon interaction, and are relatively well understood within the unitary chiral models \cite{Kaiser:1995eg,angels,ollerulf,Lutz:2001yb,Oset:2001cn,Hyodo:2002pk,Jido:2003cb,Borasoy:2005ie,Oller:2006jw,Borasoy:2006sr,Nieves:1999bx,Garcia-Recio:2003ks}. The low lying $J^P= 1/2^+$ domain remains far less understood, both experimentally and theoretically. For instance, quark models seem to face difficulties in reproducing properties of the resonances in this sector \cite{Glozman:1995fu}. The neat reproduction of the low lying $1/2^-$ states in the $S$-wave meson-baryon interaction, using chiral dynamics, suggests that the addition of a pseudoscalar meson in S-wave could lead to an important component of the structure of the $1/2^+$ resonances.  Chiral dynamics has been used earlier in the context of the three nucleon problems, e.g., in \cite{epelbaum}. We present here the study done in \cite{MartinezTorres:2007sr} of two meson - one baryon systems, where chiral dynamics is applied to solve the Faddeev equations. As described below, our calculations for the total $S$ = -1 reveal peaks in the  $\pi \bar{K} N$ system and its coupled channels which can be identified with the resonances $\Sigma(1770)$, $\Sigma(1660)$, $\Sigma(1620)$, $\Sigma(1560)$, $\Lambda(1810)$ and $\Lambda(1600)$.

\section{The formalism for three body systems}
We start by taking all combinations of a pseudoscalar meson of the $0^-$ SU(3) octet and a baryon of the $1/2^+$ octet which couple to $S=-1$ with any charge. For some quantum numbers, the interaction of this two body system is strongly attractive and responsible for the generation of the two $\Lambda(1405)$ states \cite{Jido:2003cb} and other $S$ = -1 resonances. We shall assume that this two body system formed by $\bar{K}N$ and coupled channels remains highly correlated when a third particle is added, in the present case a pion. Yet, the formalism allows for excitation of this cluster in intermediate steps.  Altogether, we get twenty-two coupled channels for the net charge zero configuration: $\pi^0 K^- p$, $\pi^0\bar{K}^0 n$, $\pi^0\pi^0\Sigma^0$, $\pi^0\pi^+\Sigma^-$, $\pi^0\pi^-\Sigma^+$, $\pi^0\pi^0\Lambda$, $\pi^0\eta\Sigma^0$, $\pi^0\eta\Lambda$, $\pi^0 K^+\Xi^-$, $\pi^0 K^0\Xi^0$, $\pi^+ K^- n$, $\pi^+\pi^0\Sigma^-$, $\pi^+\pi^-\Sigma^0$, $\pi^+\pi^-\Lambda$, $\pi^+\eta\Sigma^-$, $\pi^+ K^0\Xi^-$, $\pi^-\bar{K}^0 p$, $\pi^-\pi^0\Sigma^+$, $\pi^-\pi^+\Sigma^0$, $\pi^-\pi^+\Lambda$, 
$\pi^-\eta\Sigma^+$, $\pi^- K^+ \Xi^0$. We assume the correlated pair to have a certain invariant mass, $\sqrt{s_{23}}$, and the three body $T$-matrix is evaluated as a function of  this mass and the total energy of the three body system. At the end we look for the value of $|T|^2$ as a function of these two variables and find peaks at certain values of these two variables, which indicate the mass of the resonances and how a pair of particles is correlated.  

The input requires to solve the Faddeev equations, i.e., the two body $t$-matrices for the meson-meson and meson-baryon interactions is calculated by taking the lowest order chiral Lagrangian following
\cite{npa,angels,Oset:2001cn,Inoue} and using the dimensional regularization of the loops as done in
\cite{ollerulf,Oset:2001cn}, where a good reproduction of scattering amplitudes and resonance properties was found. Improvements introducing higher order Lagrangians have been done recently \cite{Oller:2006jw,Borasoy:2005ie,Borasoy:2006sr}, including a theoretical error analysis in \cite{Borasoy:2006sr} which allows one to see that the results with the lowest order Lagrangian fit perfectly within the theoretical allowed bands.

A shared feature of the recent unitary chiral dynamical calculations is the on-shell factorization of the potential and the $t$-matrix in the Bethe-Salpeter equation \cite{angels,ollerulf,Nieves:1999bx,Garcia-Recio:2003ks,Hyodo:2002pk,Borasoy:2005ie,npa}, which is
justified by the use of the N/D method and dispersion relations \cite{nsd,ollerulf}. Alternatively, one can see that the off-shell contributions can be reabsorbed into renormalization of the lower order terms \cite{npa,angels}. We develop here a similar approach for the Faddeev equations.

The full three-body $T$-matrix can be written as a sum of the auxiliary $T$-matrices $T^1$, $T^2$ and $T^3$ \cite{Faddeev}
\begin{equation}
T=T^1+T^2+T^3
\end{equation}
where $T^i$, $i=1$, $2$, $3$, are the normal Faddeev partitions, which include all the possible interactions contributing to the three-body $T$-matrix with the particle $i$ being a spectator in the last interaction.
The Faddeev partitions satisfy the equations
\begin{equation}\label{eq:Tiorig}
T^i=t^i\delta^3(\vec{k}^{\,\prime}_i-\vec{k}_i)+ t^i g^{ij}T^j + t^i g^{ik}T^k ,
\end{equation}
where $\vec{k}_i$ ($\vec{k}^{\,\prime}_i$) is the initial (final) momentum of the ith particle in the global center of mass system, $t^i$ is the two-body $t$-matrix for the interaction of the 
pair $(jk)$ and $g^{ij}$ is the three-body propagator or Green's function, with $j \neq k \neq i$ = 1, 2, 3

\begin{figure}[ht]
\includegraphics[width=0.6\textwidth] {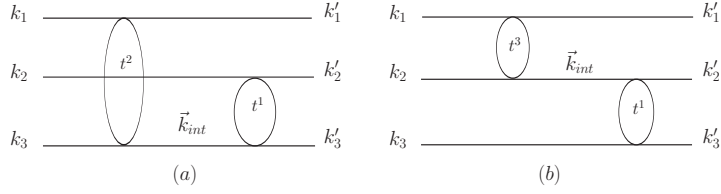}
\caption{\label{fig1} Diagrammatic representation of the terms (a) $t^1 g^{12} t^2$
(b) $t^1 g^{(13)} t^3$.}
\end{figure}

\begin{figure}[ht]
\includegraphics[scale=0.6] {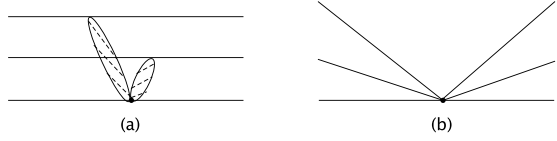}
\caption{\label{fig2} The origin of the three body forces (a) due to cancellation of the propagator in Fig.\ref{fig1}(a) with the off-shell part of the chiral amplitude, (b) at the tree level from the chiral Lagrangian.}
\end{figure}

The first two terms of the diagrammatic expansion of the Faddeev equations, for the case $i$=1, are represented diagrammatically in Fig.\ref{fig1},
where the $t$-matrices are required to be off-shell. However, the chiral amplitudes, which we use,
can be split into an ``on-shell'' part (obtained when the only propagating particle of the diagrams, labeled with $\vec{k}_{int}$ in Fig.\ref{fig1}, is placed on-shell (meaning that $\vec{k}^2_{int}$ is replaced by $m^2$ in the amplitudes),  and an off-shell part proportional to the inverse of the propagator of the off-shell particle, $\vec{k}^2_{int}-m^2$. This term would cancel the particle propagator, ($\vec{k}^2_{int}-m^2)^{-1}$, for example that of the 3rd particle in Fig.\ref{fig1}a) resulting into a three body force (Fig.\ref{fig2}a). In addition to this,
three body forces also stem directly from the chiral Lagrangians \cite{Felipe} (Fig.\ref{fig2}b).

We find that the sum of the off-shell parts of all the two interaction terms of the Faddeev series, cancel together with the contribution from Fig.\ref{fig2}(b) in the SU(3) limit. Details of the analytical proof can be seen in the appendices of \cite{MartinezTorres:2007rz,MartinezTorres:2008gy}. Hence, only the on-shell part of the two body (chiral) $t$-matrices is needed in the evaluations. This is one of the important findings of these works because one of the standing problems of the Faddeev equations is that the use of different potentials which give rise to the same on shell scattering amplitudes give rise to different results when used to study three body systems with the Faddeev equations. The different, unphysical, off shell amplitudes of the different potentials are responsible for it. The use of chiral dynamics in the context of the Faddeev equations has then served to show that the results do not depend on these unphysical amplitudes and only the on shell amplitudes are needed as input. In this sense, since these amplitudes can be obtained from experiment, it is suggested in \cite{MartinezTorres:2008kh} to use these experimental amplitudes, and sensible results are obtained in the study of the $\pi \pi N $ system and coupled channels.

 The strategy followed in the former works is that the terms with two, three, and four interactions are evaluated exactly. Then it is observed that the ratio of the four to three body interaction terms is about the same as that of the three body to two body. Once this is realized, the coupled integral equations are converted into algebraic equations, which renders the technical work feasible in spite of the many coupled channels used.

The resonances generated for these systems appear as peaks in $|T|^2$ as a function of $\sqrt{s}$, $\sqrt{s_{23}}$.
A detailed description of all the states that appear in this sector can be seen in \cite{MartinezTorres:2007sr}. Here we summarize the results in Table \ref{table1}. One should note that, quite systematically, the widths obtained are smaller then experiment. This is due to the neglect of the one meson-baryon channels. The idea is that they have small influence in the wave function, which is largely three-body, but they still contribute to the width because there is far more phase space for the decay.

\begin{table}
\centering
\begin{tabular}{cccc}
\hline
&$\Gamma$ (PDG)&Peak position& $\Gamma$ (this work)\\
&(MeV)&(this work, MeV)& (MeV)\\
\hline
Isospin=1&&&\\
\hline
$\Sigma(1560)$&10-100&1590&70\\
$\Sigma(1620)$&10-100&1630&39\\
$\Sigma(1660)$&40-200&1656&30\\
$\Sigma(1770)$&60-100&1790&24\\
\hline
Isospin=0&&&\\
\hline
$\Lambda(1600)$&50-250&1568,1700&60-136\\
$\Lambda(1810)$&50-250&1740&20\\
\hline

\end{tabular}
\caption{$\Sigma$ and $\Lambda$ states obtained from the interaction of two mesons and one baryon.}\label{table1}
\end{table}

  In the S=0 sector we also find several resonances, which are summarized in Table \ref{table2}. Since this work is mostly about strangeness, we only want to pay attention to the $N^*$ state around 1924 MeV, which is mostly $N K \bar{K}$. This state was first predicted in \cite{Jido:2008kp} using variational methods and corroborated in \cite{MartinezTorres:2008kh} using coupled channels Faddeev equations. As in \cite{Jido:2008kp}, we find that the $ K \bar{K}$ pair is built mostly around the $f_0(980)$, but it also has a similar strength around the $a_0(980)$, both of which appear basically as a $ K \bar{K}$ molecule in the chiral unitary approach.
  
\begin{table}
\centering
\begin{tabular}{c|ccc|ccc}
\hline\hline
$I(J^P)$&\multicolumn{3}{c}{Theory}&\multicolumn{3}{c}{PDG data}\\
\hline
&channels&mass&width&name&mass&width\\
&&(MeV)&(MeV)&&(MeV)&(MeV)\\
\hline
$1/2(1/2^+)$&only $\pi\pi N$&1704&375&$N^*(1710)$&1680-1740&90-500\\
& $\pi\pi N$, $\pi K\Sigma$, $\pi K\Lambda$, $\pi\eta N$&$\sim$ no change&$\sim$ no change&&&\\
\hline
$1/2(1/2^+)$&only $\pi\pi N$&2100&250&$N^*(2100)$&1885-2270&80-400\\
& $\pi\pi N$, $\pi K\Sigma$, $\pi K\Lambda$, $\pi\eta N$&2080&54&&&\\
\hline
$3/2(1/2^+)$&$\pi\pi N$, $\pi K\Sigma$, $\pi K\Lambda$, $\pi\eta N$&2126&42&$\Delta(1910)$&1870-2152&190-270\\
\hline
$1/2(1/2^+)$&$\pi\pi N $, $\pi\eta N$, $K\bar K N$&1924&20&$N^*(?)$&?&?\\
\hline\hline
\end{tabular}
\caption{$N^{*}$ and $\Delta$ states obtained from the interaction of two mesons and one baryon.}\label{table2}
\end{table}
  
  This state is very interesting and it was suggested in  \cite{conulf} that it could be responsible for the peak around 1920 MeV of the $\gamma p \to K^+ \Lambda$ reaction \cite{saphir,jefflab,mizuki}. It was also shown that the spin of the resonance could be found performing polarization measurements which are the state of the art presently.

\section{The $\bar{K}NN$ system}

 The two Faddeev approaches \cite{Ikeda:2007nz,Shevchenko:2006xy} lead to binding energies higher than the variational approach, 50-70 MeV versus around 20 MeV binding
 respectively. The detail mentioned above of an energy independent kernel used in the AGS equations is partly responsible for the extra
 binding of these approaches with respect to the chiral calculations. Indeed, 
 the chiral potential is energy dependent, proportional to the sum of the two external meson energies in $\bar{K}N \to \bar{K}N$.
 As a consequence, a smaller  $\bar{K}N \to \bar{K}N$ amplitude is obtained at lower $\bar{K} N$ energies, resulting in a smaller binding 
for the $\bar{K}NN$ system. This numerical result was already mentioned in \cite{Dote:2008hw}. Actually, the same result is found within 
the approach of \cite{Ikeda:2007nz,Ikeda:2008ub} when the energy dependence of the Weinberg-Tomozawa chiral potential is taken into account
 in \cite{Ikeda:2010tk}.
 An important step to conjugate the AGS equations
with the $\bar{K} N-\pi \Sigma$ dynamics of the chiral Lagrangians has been done, in \cite{Ikeda:2010tk}. Indeed, two poles
are found in qualitative agreement with other chiral approaches. One narrow pole around 1420 MeV, rather stable against changes of parameters,
is found in agreement with all findings of the chiral unitary approach. The second, wider pole, is found at very low energies 1335-1341 MeV, and 
more unstable with respect to changes of parameters. This agrees qualitatively with the findings of the chiral unitary approach, but the energy
is lower than in other approaches. One point to try to understand these difference is that in \cite{Ikeda:2010tk} the $\pi \Sigma$ mass 
distribution of the $K^- p\rightarrow\pi\pi\pi\Sigma$ in the Hemingway experiment \cite{Hemingway:1984pz} is adjusted assuming that it is proportional
to $|T_{\pi\Sigma,\pi\Sigma}|^{2}$. Yet, as shown in \cite{Jido:2003cb}, when one has two poles, the $T_{\pi\Sigma,\pi\Sigma}$
 and $T_{\bar{K} N,\pi\Sigma}$
amplitudes are rather different and the $\Lambda(1405)$ production processes proceed via the combination of the two amplitudes
$|T_{\pi\Sigma,\pi\Sigma}+ \beta ~T_{\bar{K} N,\pi\Sigma}|^{2}$ \cite{Hyodo:2003jw}.

Yet, in most cases the widths are 
systematically larger than the binding energy, of the order of 70-100 MeV. This certainly makes the observation of these states problematic, 
as acknowledged in all these works. In view of this, the claim in \cite{Agnello:2005qj} of a bound state of $K^- pp$ bound state with 115 MeV
 binding was met with skepticism, and soon it  was shown that the peak observed in 
\cite{Agnello:2005qj} was easily interpreted in terms of conventional, unavoidable, reaction mechanisms which were well under control 
\cite{Magas:2006fn}.
     The same was done with the claim of a $K^- NN$ cluster in \cite{:2007ph} which was also dismissed in \cite{Magas:2008bp} on the grounds that a conventional explanation could be found for it.

   In Ref. \cite{melahat} the authors follow the chiral unitary approach for the $\bar{K}N $ amplitudes, which provide the most 
important source of the binding of the three body system, according to the former studies. But there is also another different technical 
aspect of that calculation with respect to the former ones. All previous approaches have concentrated on looking for the binding,
 searching for poles in the complex plane or looking for the energy that minimizes the expectation value of the Hamiltonian.  There, inspired 
by the studies of   
\cite{MartinezTorres:2007sr,Khemchandani:2008rk,MartinezTorres:2008gy}, One looks for peaks in the scattering matrices as a function of
 the energy of the three body system. These amplitudes  could in principle be used as input for final state 
interaction when evaluating cross sections in reactions where eventually this  $\bar{K}N N$ state is formed.

\section{Fixed Center formalism for the $\bar{K} NN$ ($S=1$) interaction}

  The findings of the former works simplify our task. We rely upon the results of 
\cite{MartinezTorres:2007sr,Khemchandani:2008rk,MartinezTorres:2008gy} in the sense that only on shell two body amplitudes are needed. By this we mean the part of the analytical amplitudes obtained setting $q^2=m^2$ for the external particles, even if the particles are below threshold.

We also assume, like in the other works, that the two body interactions proceed in L=0. 
  According to all the works, the main component of the wave function corresponds to having a $\bar{K} N $ in I=0 and hence the total isospin will be I=1/2. The total spin can be J=0,1, but the  $J^P=0^-$ state is the one found most attractive. Both possibilities are investigated in \cite{melahat}.

   In the FCA the $\bar{K}NN$ system is addressed by studying  the interaction of a $\bar{K}$ with a $NN$ cluster.
 The scattering matrix for this system is evaluated as a function of the total energy of the  $\bar{K}NN$ system and one looks for peaks in $|T|^2$.
In a second step one allows explicitly the intermediate  $\pi\Sigma N$ state in the three body system.
 The FCA has been used before in connection with the evaluation $K^{-}d$  
scattering length  \cite{Toker:1981zh,Chand:1962ec,Barrett:1999cw,Deloff:1999gc,Kamalov:2000iy}. A follow up of \cite{Kamalov:2000iy}
is done in \cite{Meissner:2006gx}. A discussion of these different approaches is done in \cite{Gal:2006cw}, where it is shown that Refs. \cite{Barrett:1999cw, Deloff:1999gc}
do not take into account explicitly the charge exchange $K^{-}p\rightarrow\bar{K}^{0}n$ reactions and antisymmetry of the nucleons, while 
it is explicitly done in Ref. \cite{Kamalov:2000iy}.

The FCA for $K^{-}d$ at threshold was found to be an acceptable approximation, within 20-25$\%$ percent, to the more elaborate Faddeev equations
in \cite{Toker:1981zh,Gal:2006cw}. 
Technically we follow closely the formalism of \cite{multirho}, where the FCA has been considered, using chiral amplitudes, in order to
 study theoretically the possibility of forming multi-$\rho$ states with large spin.

An interesting shared result of all the $K^{-}d$ calculation quoted above is a large and negative real part of the scattering length, of the order 
of $-1.40+i 1.80$ fm, which suggest the existence of a bound state in $ J=1$. Of course the imaginary part of the scattering length is equally
large, anticipating a broad state. We shall be able to make this more quantitative here.

The strategy followed in \cite{melahat} is to assume as a starting point that the $NN$ cluster has a wave function like the one of the deuteron (we  omit the d-wave).
Latter on one releases this assumption and assumes that the $NN$ system is further compressed in the ${\bar K}NN$ system.
Since in the FCA the input from the $NN$ system is the $NN$ form factor, taking into account an extra compression of the $NN$ systems is very easy by smoothly modifying the form factor to have a smaller radius.
One takes information from previous studies, and in this sense it is interesting to recall that the calculations of \cite{Mares:2006vk} for ${\bar K}$ bound in nuclei point out to a moderate compression of the nucleus due to the strong ${\bar K}N$ interaction.
However, in the case of two isolated nucleons the decrease in the radius can be far bigger that in nuclei, where nucleons are already
close to saturation density. The information on the $NN$ radius in the ${\bar K}NN$ molecule we get from \cite{Dote:2008hw}, where the $NN$
interaction is taken into account including short range repulsion that precludes unreasonable compression. The r.m.s radius found is of 
the order of 2.2 fm, slightly above one half the value of the deuteron r.m.s radius of 3.98 fm ($NN$ distance).

 The formal derivation follows the steps of \cite{multirho}.
We assume pure $L=0$ interaction for all the ${\bar K}N$ and $NN$ pairs, and select $S=J=1$, $I_{NN}=0$ for the two nucleons. Since $L=0$ also for ${\bar K}N$, $J$ will be the spin of the total ${\bar K}NN$ system and 
 the total isospin will be $I=1/2$.

\section{Consideration of charge exchange steps } 

The formula of Eq. (19) in Ref. \cite{melahat} does not allow for intermediate charge exchange processes, $K^- p \rightarrow \bar{K}^0 n$ followed by $\bar{K}^0 n \rightarrow K^- p$. These were explicitly taken into account in \cite{Kamalov:2000iy} and found to be relevant in $K^-d$ scattering. In this section we take into account this extra possible step and see how the results change. The right expression considering charge exchange processes can be found from Eq. (24) of Ref. \cite{Kamalov:2000iy} substituting $\tilde a$ by the t matrix and $1/r \rightarrow G_0$. Thus, we obtain for $K^- d$ the amplitude 
\begin{figure}
\begin{center}
\centerline{\includegraphics[width=9.5cm]{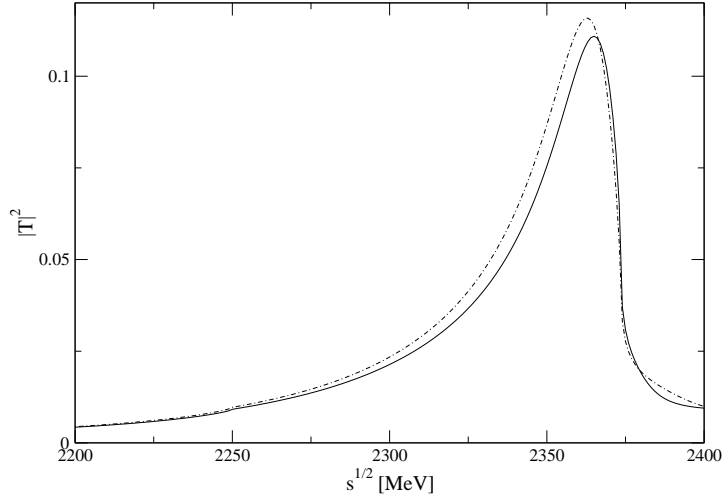}}
\caption{Modulus squared of the three-body scattering amplitude for triplet (S=1).  The solid line indicates the result of Ref. \cite{Kamalov:2000iy} with normal size. The dot-dashed line indicates the result of  Ref. \cite{Kamalov:2000iy} with the reduced radius.\label{fig:tyeni}}
\end{center}
\end{figure}

\begin{equation}
T_{K^- d}=\frac{t_p+t_n+(2t_pt_n-t_x^2)G_0-2 t_x^2t_nG_0^2}{1-t_pt_nG_0^2+t_x^2t_nG_0^3}
\label{Eq:kamalov}
\end{equation} 
where
$t_p=t_{K^- p \rightarrow K^- p}$, $t_n=t_{K^- n \rightarrow K^- n}$, $t_x=t_{K^- p \rightarrow \bar{K}^0 n}/ \sqrt{1+t_n^0 G_0}$, $t_n^0=t_{\bar{K}^0 n \rightarrow \bar{K}^0 n}$,
and
\begin{equation}
G_0=\int\frac{d^3q}{(2\pi)^3}F_{NN}(q)\frac{1}{{q^0}^2-\vec{q}\,^2-m_{\bar K}^2+i\epsilon}.
\label{Eq:kamalov1}
\end{equation}
In principle one must also consider the form factor in the single scattering \cite{multirho}, $F((\vec k- \vec k')/2)$, with $\vec k,~ \vec k'$  the initial and final $\bar{K}$ momenta, which is 1 for $\vec k= \vec k'$. For the typical momenta of  $\vec k$ in this bound state this factor does not differ much from 1 in any case and we take it equal 1, as also done in \cite{multirho}. The variable $q^0$ in Eq.~(\ref{Eq:kamalov1}) is the energy carried by the ${\bar K}$. For the form factor we take, as a starting point, the one of the deuteron. In further steps this form factor will be changed to accommodate the reduced size of 
the two $N$ system found in \cite{Dote:2008hw}.

By using isospin symmetry one can see that the amplitudes can be written as 
\begin{equation}
t=b^0+b^1 \vec\tau_{\bar{K}}.\vec\tau_N;~~ b_0=\frac{1}{4}(3t^1+t^0),~~ b_1=\frac{1}{4}(t^1-t^0),
\end{equation}
where $t^0$, $t^1$ are the $\bar{K}N$ amplitudes in isospin 0,1, respectively. Following Ref. \cite{Gal:2006cw} we can write now Eq. (\ref{Eq:kamalov}) as

\begin{equation}
T_{K^- d}=\frac{2b_0-2(b_0+b_1)(3b_1-b_0)G_0}{1-2b_1G_0+(b_0+b_1)(3b_1-b_0)G_0^2}
\label{Eq:gal}
\end{equation}
and then we immediately see that if we neglect $b_1$, taking it equal zero, we regain Eq. (19) of Ref. \cite{melahat}. It is thus clear that we made the approximation of neglecting the isospin flip part of the amplitude, $b_1$ in \cite{melahat}. In view of this, we redo here the calculations using the more accurate expression of Eq. (\ref{Eq:kamalov}). The new results can be seen in Fig. \ref{fig:tyeni}. We observe, by comparison to the results omitting $b_1$ shown in \cite{melahat}, that the peak has been shifted to lower energies by about 12 MeV. We also show results with the normal and "reduced" size. A smaller size like the one found in \cite{Dote:2008hw} for $I_{NN}=1$ leads to a slightly more bound state by 2 MeV. The binding energy obtained is of 9 MeV and  the width of about 30 MeV.

\section{Explicit consideration of the $\pi \Sigma N $ channel in the three body system}

\begin{figure}[!t]
\begin{center}
\includegraphics[width=0.25\textwidth]{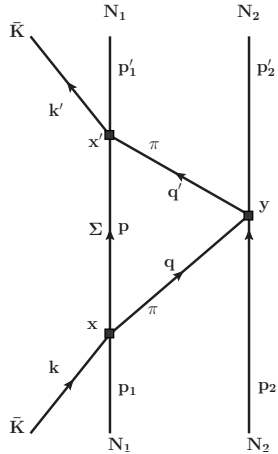}
\caption{Feynman diagram  for the $\pi \Sigma N$  channel.
The equivalent diagrams where the ${\bar K}$ interacts first with the second nucleon should be added.}
\label{fig:10}
\end{center}
\end{figure}

As we have explained, the $\pi \Sigma$ channel and other coupled channels are explicitly taken into account in the consideration of the 
$\bar{K}N$ amplitude which we have used in the FCA approach. This means that we account for the $\bar{K}N \rightarrow  \pi \Sigma$ transition,
and an intermediate  $\pi \Sigma N $ channel, but this  $\pi \Sigma $ state is again reconverted to $\bar{K}N$ leaving the other $N$
as a spectator. This is accounted in the multiple scattering in the  $\bar{K}N \rightarrow \bar{K}N $ scattering matrix on one nucleon. However, we do not consider the possibility
that one has $\bar{K}N \rightarrow  \pi \Sigma$ and the $\pi$ rescatters with the second nucleon. If we want to have a final $\bar{K}N N$
system again, the $\pi$ has to scatter later with the $\Sigma$ to produce $\bar{K}N$. One may consider multiple scatterings of the $\pi$ 
with the nucleons, but given the smallness of the $\pi N$  amplitude compared to the $\bar{K}N$, any diagram beyond the one having one rescattering
of the pion with the nucleon will be negligible. Then, to account for explicit $\pi \Sigma N$ state in the there body system one is left with the diagram  of
Fig. ~\ref{fig:10} (for $\bar{K}N$ scattering on nucleon 1). 

\begin{figure}
\begin{center}
\centerline{\includegraphics[width=9.5cm]{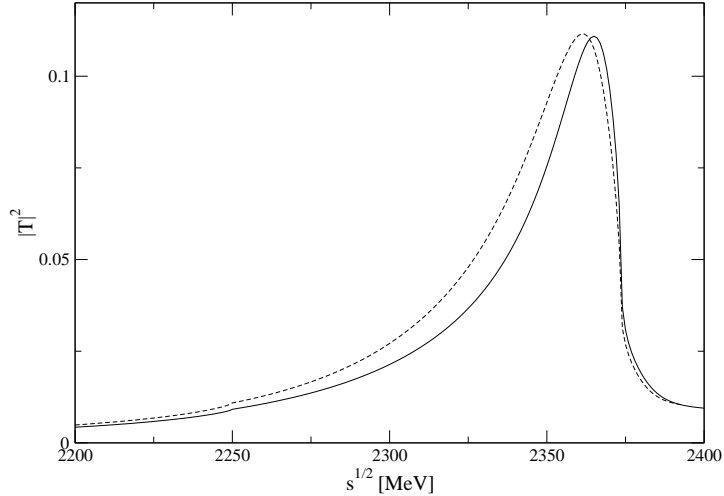}}
\caption{Modulus squared of the three-body scattering amplitude for triplet (S=1). The solid line indicates the $\bar{K}NN$ system in normal size 
and the dashed line indicates the result of including
 the $\pi \Sigma$ channel with the normal size.\label{fig:tmats1} }

\end{center}
\end{figure}

The result of including the $\pi \Sigma N$ channel can be seen in Fig.~\ref{fig:tmats1}  for $S=1$. We compare the results with $NN$ normal size with and without explicit  $\pi \Sigma N $ channel in the three body state.
The effects of the $T^{(\pi \Sigma)}$ amplitude are small and we see a change in the peak position of about 2.5 MeV to lower energies and a small increase in $|T|^{2}$ as a result of the explicit consideration of the $\pi \Sigma N$ channel. The results with the increased $NN$ size are similar.

\section{Threshold results } 

The work done also allows to get the threshold amplitude for $K^-d$, which is our $\bar{K}NN ~ (S=1)$ state. The relationship of our scattering amplitude $T$, to the ordinary one of quantum mechanics, $f(k)$, is 
 
\begin{eqnarray}
f(k)=-\frac{1}{4 \pi}\frac{M_d}{\sqrt{s}}T;~~~\lim_{k\rightarrow0} f(k)=a.
\end{eqnarray}
We obtain 

\begin{eqnarray}
\label{Eq:yyy1111}
a\simeq(-1.54+i1.82)~fm.
\end{eqnarray}

The scattering length coincides with the one of Ref. \cite{Kamalov:2000iy} of $(-1.6+i1.9)~fm$. The different regularization of the loops (dimensional regularization here and cut off in 
\cite{Kamalov:2000iy}) can be blamed for the tiny differences. These results compare with 
$(-1.47+i1.08)~fm$ from \cite{Toker:1981zh}, $(-1.34+i1.04)~fm$ from \cite{Torres:1986mr},  $(-1.80+i1.55)~fm$ from 
\cite{Bahaoui:1990da} and $(-0.78+i1.23)~fm$ from \cite{Grishina:2004rd}. This gives us an idea of the dispersion of the different results. 
	
	A recent work which combines the ${\bar K}N$ scattering data and the $K^-p$ atomic data from SIDDHARTA \cite{Bazzi:2011zj} provides the $K^-d$ scattering length $(-1.46+i1.08)~fm$ from \cite{Doring:2011xc}, but a study of uncertainties is done resulting in a broad region of allowed values where the result of Eq. (\ref{Eq:yyy1111}) fits well.
One should note that the values of $a$ mentioned above hint at a possible $K^-d$ bound state using $k \rightarrow i \kappa$, $B=\kappa^2/2 M_d $, $\kappa=-1/a$, but, since one is dealing with a resonant $\bar{K}N$ amplitude, the explicit calculation as we have done before gives a more reliable result.

\section{Conclusions} 

We present new experimental and theoretical results for the 
$K^{-} d \to \pi \Sigma n$ reaction which give strong support to the claim of two $\Lambda(1405)$ states of the chiral unitary theories.  On the other hand we have reported a new approach to Faddeev equations based on chiral unitary dynamics in coupled channels that explicitly shows the cancellation of the off shell two body scattering matrix with three body terms stemming from the same theory. The elimination of the unphysical off shell parts renders the approach more accurate. Finally we presented results of calculations for the scattering amplitudes of the $\bar{K}NN$ system using the FCA of 
Faddeev equations considering the scattering of the light $\bar{K}$ against the heavier NN cluster. 
We assumed that a NN cluster is made, since in $S=1$, $I_{NN}=0$ the $NN$ system is bound, and the  $S=0$, 
$I_{NN}=1$, which is nearly bound in free space, gets the small push needed to bind from the strong attractive
$\bar{K}N$ interaction. 
The new size of  the $NN$ system is taken from the results for the  $NN$
radius obtained in the calculations of \cite{Dote:2008hw}. We found that the consideration of this reduced $NN$ size reverted into  
a larger binding of the three body system. We have presented here the results for $K^- d$ bound state in the Fixed center approximation considering charge exchange process, thus, improving on our previous result in Ref. \cite{melahat}. We found a bound state with 9 MeV binding and about 30 MeV width.
 This state was not searched for in the Faddeev and variational calculations which concentrated on the $S=0$, $I_{NN}=1$ state, which is more bound. In view of that full Faddeev and variational calculations of this $S=1$ state should be most welcome.

  The simplicity of the present approach also allows for a transparent interpretation of the results,
not easy to see when one uses either a variational method or the full Faddeev equations. 
The dominance of the $S=0$, $I_{NN}=1$ channel could be anticipated once the $\bar{K}N$
amplitudes for a $N$ belonging to a cluster with $S=0$ or $S=1$ is known. This conclusion is in agreement 
with results of other methods, which were found with more laborious ways.

  \section{Acknowledgments}
  We thank  A. Gal and T. Hyodo  for useful comments. This work is partly supported by  projects FIS2006-03438 from the Ministerio de Ciencia e Innovaci\'on (Spain), FEDER funds and by the Generalitat Valenciana in the program Prometeo/2009/090.
This research is part of the European
 Community-Research Infrastructure Integrating Activity ``Study of
 Strongly Interacting Matter'' (acronym HadronPhysics2, Grant
 Agreement n. 227431) 
 under the Seventh Framework Programme of EU. M. Bayar acknowledges support through the Scientific and
 Technical Research Council (TUBITAK) BIDEP-2219 grant.


\end{document}